\documentclass[showpacs,aps,pre,numerical,twocolumn,10pt,superscriptaddress]{revtex4-1}
\pdfoutput=1

\usepackage{graphicx}
\usepackage{amsmath,amssymb}
\usepackage{times,mathptmx}
\usepackage{units}
\usepackage{epsfig}
\usepackage{booktabs}
\usepackage{multirow}

\usepackage[latin1]{inputenc}
\usepackage[english]{babel}
\usepackage{color}

\newcommand{\vecu}{\mathbf}

\begin{document}

\title{Failure of the $\alpha$-factor in describing QD laser dynamical instabilities and chaos}

\author{Benjamin Lingnau}\affiliation{Institut f{\"u}r Theoretische Physik, Technische Universit{\"a}t Berlin, 10623 Berlin, Germany}
\author{Kathy L{\"u}dge}\affiliation{Institut f{\"u}r Theoretische Physik, Technische Universit{\"a}t Berlin, 10623 Berlin, Germany}
\author{Weng W. Chow}\affiliation{Sandia National Laboratories, Albuquerque, New Mexico 87185-1086, USA}
\author{Eckehard Sch{\"o}ll}\affiliation{Institut f{\"u}r Theoretische Physik, Technische Universit{\"a}t Berlin, 10623 Berlin, Germany}
 
\begin{abstract}
  We show that the long-established concept of a linewidth-enhancement factor $\alpha$ to describe carrier-induced refractive index changes in semiconductor lasers breaks down in quantum dot (QD) lasers when describing complex dynamic scenarios, found for example under high-excitation or optical injection. By comparing laser simulations using a constant $\alpha$-factor with results from a more complex non-equilibrium model that separately treats gain and refractive index dynamics, we examine the conditions under which an approximation of the amplitude-phase coupling by an $\alpha$-factor becomes invalid. The investigations show that while a quasi-equilibrium approach for conventional quantum well lasers is valid over a reasonable parameter range, allowing one to introduce an $\alpha$-factor {as a constant parameter, the concept is in general not applicable 
to predict QD laser dynamics} due to the different timescales of the involved scattering processes.
\end{abstract}


\maketitle

{Semiconductor lasers are essential components in optoelectronics that affect
practically every aspect of our daily life.  There are amble indications
that the present technology is approaching a stage where fundamental constraints
are limiting performance.  The semiconductor quantum-dot (QD) laser is a strong candidate for introducing improvements at the underlying physics level.  A very important consideration is dynamical performance, which impacts
almost all applications.

In semiconductor lasers, the connection between refractive index and optical
gain plays an important role in determining modulation response and
dynamical instabilities. Throughout literature, this connection is commonly
described by assuming a linear relationship between changes of the real
and imaginary parts of optical susceptibility $\chi$, with the proportionality
given by a constant linewidth-enhancement factor $\alpha$ \cite{HEN82}.
For example, with optical injection \cite{WIE05} or feedback \cite{OTT10},
laser dynamics depends critically on the phase dynamics of the intracavity
laser field. It is customary to describe this dynamics by introducing a constant
$\alpha$ into the field equation \cite{LAN80b,SIM94a,GOU07}.

For QD lasers, the concept of an $\alpha$-factor has been controversially
discussed \cite{MEL06,GIO07}. Experimental values range from near zero
\cite{GHO02,FAT05,MI05,KIM10e} to larger values\cite{UKH04} up to $60$ \cite{DAG05,GRI08}, 
with different measurement techniques yielding very different results \cite{MEL06,GIO06a,CON07}.
Furthermore, frequency chirp under large-signal modulation was found
to be inaccurately described by $\alpha$ \cite{GIO07}.}

Nevertheless, the failure of $\alpha$ in accurately describing QD lasers is either not widely known, or ignored.
Studies of QD laser dynamics have relied on $\alpha$ to describe the amplitude-phase correlation \cite{HUY04,GRI08a,KIM10a,KEL11a,ZHU12} and experimental measurements of the carrier-induced refractive index in QD active media have been expressed in terms of the $\alpha$-factor \cite{KIM06,CON07,ZIL08,GER08}.
In this letter, we discuss situations where the use of the $\alpha$-factor leads to incorrect predictions. In particular, for the cases involving optical injection or feedback, which are widely studied experimentally and theoretically by the laser dynamics community \cite{AZO07,ERZ07a,NAD09,KEL09,ERN10a,POC11,SCH11d,FIO11,PAU12,OTT12}, we present examples where very different laser dynamics is predicted if an $\alpha$-factor is used instead of the full microscopic description.

Our model for the QD laser device includes the full time dependence of the polarization of the active medium and allows us to derive the gain and the refractive index in each time step without the need to introduce an $\alpha$-factor. We apply a semi-classical approach using Maxwell's equations and the semiconductor-Bloch equations \cite{CHO99,LIN12a}. 
Carriers are injected from the bulk into a carrier reservoir (QW), and from there into the QDs. A bound ground state (GS) and a twofold degenerate (excluding spin degeneracy) excited state (ES) of the QDs are considered, both for electrons and holes. The bound states are labeled by the index $m\in\{GS,ES\}$. 
In order to account for the inhomogeneous broadening of the QD transitions, we distribute the QDs into different subgroups, labeled by an index $j$, with the transition frequency $\omega_m^j$. The distribution function $f(j)$ gives the probability to find a specific QD in the $j$-th subgroup, such that $\sum_jf(j)=1$. We assume a Gaussian spectral distribution of the QDs with a width (FWHM) of $60$meV. 
The following dynamic equations describe the time evolution of the slowly varying electric field $E(t)$ and the occupation probabilities of the QD subgroups $\rho^j_{b,m}$, as well as the QW and bulk carrier densities per unit area $n^\mathrm{QW}_b$ and $n^\mathrm{bulk}_b$, respectively ($b\in\{e,h\}$ distinguishes between electrons and holes):
\begin{align}
  \frac{d}{dt} E &= g(\omega,t) E - \kappa E + K \frac{E^0}{\tau_{in}} \exp[-i(\omega_{inj}-\omega)t] \label{Edyn}\\
  \frac{d}{dt} \rho^j_{b,m}  &= \frac{1}{\hbar}\mathrm{Im}\big(p^j_m\mu_m^*E^*\big)-W_m\rho^j_{e,m}\rho^j_{h,m} + \frac{\partial}{\partial t} \rho^j_{b,m} \Big|_{col} \label{QDcarriereq}\\
  \frac{d}{dt} n^\mathrm{QW}_b  &= \frac{2}{A}\sum_{k^{2D}}\frac{1}{\hbar}\mathrm{Im}\big(p^{2D}_k\mu_{QW}^*E^*\big) - B^S n^\mathrm{QW}_e n^\mathrm{QW}_h  + \frac{\partial}{\partial t} n^\mathrm{QW}_b \Big|_{col} \label{QWcarriereq}\\
  \frac{d}{dt} n^\mathrm{bulk}_b  &= \frac{J}{e_0} - B^{S}_{bulk} n^\mathrm{bulk}_e n^\mathrm{bulk}_h + \frac{\partial}{\partial t} n^\mathrm{bulk}_b \Big|_{col} \label{bulkcarriereq}
\end{align}
Here, $\kappa$ is the optical cavity loss rate, $K$ is the injection strength of the external optical signal, scaled by the free-running electric field amplitude $E^0$ and the cavity round-trip time $\tau_{in}$. The injected signal is assumed to be a monochromatic wave with the frequency $\omega_{inj}$. The Einstein coefficient $W_m$ gives the spontaneous recombination rate in the QDs. The charge carrier losses in the QW and in the bulk are given by a bimolecular recombination rate $B^S$ and $B^S_{bulk}$, respectively. The sum in Eq.(\ref{QWcarriereq}) describes the charge carrier losses in 
the QW due to stimulated recombination between states described by an in-plane vector $k^{2D}$ (the factor of two accounts for spin degeneracy). $A$ is the in-plane device area. The pump current density is given by $J$, with the electron charge $e_0$.
The complex amplitude gain $g(\omega, t)$ is calculated from the adiabatically eliminated microscopic polarization amplitudes of the QD and QW transitions $p^j_m$ and $p^{2D}_k$, respectively \cite{GIO07,LIN12a}:
\begin{align}
  g(\omega, t) \!&=\! \frac{i\omega\Gamma}{2\varepsilon_{bg}\varepsilon_0}\frac{2{N^{QD}}}{h^{QW}}\!\left[\!\sum_{m,j} f(j) \nu_m \mu_m^* \frac{p^j_m(t)}{E(t)}\!\!\right] \!+\! \mathrm{Re}(g^{QW})\,\label{amplitudegain}\\
  p_m^j(t) &= -\frac{\mu_m T_2}{2\hbar}(\rho^j_{e,m} + \rho^j_{h,m} - 1) \frac{\Delta\omega^j_mT_2+i}{1+(\Delta\omega^j_m T_2)^2} E(t)\,\label{pol_adiab}
\end{align}
where $\mu_m$ is the respective QD dipole transition moment. $\Gamma$ is the geometric confinement factor, $\varepsilon_{bg}$ the background permittivity, $h^{QW}$ the height of each QW layer, $N^{QD}$ the QD areal density per QW layer, and $\nu_m$ the degree of degeneracy of the confined QD states. The transition frequency detuning is $\Delta\omega_m^j \equiv (\omega_m^j-\omega)$ and the polarization dephasing time constant is $T_2$. 
Here, the QW transitions are assumed to be detuned far enough from the QD GS transition to not influence the gain appreciably, such that only index changes due to the QW are taken into account, with $g^{QW}=(i\omega\Gamma)/(\varepsilon_{bg}\varepsilon_0 A h^{QW})\big[\sum_{\vecu k}\mu_{QW}p^{2D}_k(t)/E(t)\big]$. For the polarization amplitudes $p^{2D}_k$ of the QW transitions, an expression similar to Eq.~\eqref{pol_adiab} is used, with the corresponding dipole moment $\mu_{QW}$ and transition frequency detuning of the QW transitions. 
The real and imaginary parts of $g(\omega,t)$ correspond to the electric field amplitude gain and frequency shift, respectively.

The collision terms $\partial/\partial t\big|_{col}$ in Eqs.~\eqref{QDcarriereq}-\eqref{bulkcarriereq} describe the scattering between the different carrier states in the device. 
We account for direct carrier (Auger-) capture processes from the QW into the QD GS and ES, as well as relaxation processes between ES and GS, with microscopically calculated scattering rates that depend nonlinearly on the QW carrier densities \cite{MAJ10}. 
Additionally, we consider carrier-carrier ($c$-$c$) and carrier-phonon ($c$-$p$) scattering between QW and bulk states, described within the relaxation rate approximation \cite{CHO05}, with relaxation rates of $20$ps$^{-1}$ ($c$-$c$) and $4$ps$^{-1}$ ($c$-$p$) (hole rates twice as fast).

To compare with the QD laser we also model a QW laser device by taking only the QW and bulk charge carrier subsystems into account, which is obtained by eliminating Eq.~\eqref{QDcarriereq} from the dynamic equations. In order to describe nonequilibrium distributions in the QW, we model the $k$-resolved carrier distribution. The amplitude gain is then given by $g^{QW}$.

\begin{table}[tb]
\centering
{\small
\noindent\begin{tabular}{llllllll}
\hline\toprule
symbol~~ & ~~value & ~~~&symbol~ & ~~~value &~~ &symbol~~&value \\
\cline{1-2}\cline{4-5}\cline{7-8}
$\kappa$  &  $0.05$\,ps$^{-1}$
&& $\tau_{in}$  &  $48$\,ps
&& $T_2$  &  $100$\,fs
\\\addlinespace
$\mu_{GS}$  &  $0.60\,e_0$nm
&& $\mu_{ES}$  &  $0.40\,e_0$nm
&& $\mu_{QW}$  &  $0.50\,e_0$nm
\\\addlinespace
$W_{GS}$  &  $0.44$\,ns$^{-1}$
&& $W_{ES}$  &  $0.24$\,ns$^{-1}$
&& $B^S$(QD)  &  $540$\,ns$^{-1}$nm$^{2}$
\\\addlinespace
$\Gamma$  &  $0.15$
&& $\hbar\omega$  &  $0.952$\,eV
&& $B^S$(QW)  &  $54$\,ns$^{-1}$nm$^{2}$
\\\addlinespace
$\varepsilon_{bg}$  &  $14.2$
&& $h^{QW}$  &  $4$\,nm
&& $N^{QD}$  &  $10^{11}$\,cm$^{-2}$
\\[2pt]\hline\bottomrule
\end{tabular}\caption{Numerical parameters used in the simulation.}\label{tab}
}
\end{table}

In general, the $\alpha$-factor is defined as the ratio of the derivatives of the real and imaginary part of the optical susceptibility $\chi(\omega)$ with respect to the total charge carrier number $N=\sum_b\left(n^\mathrm{QW}_b+n^\mathrm{bulk}_b+2N^{QD}\sum_{j,m}f(j)\nu_m\rho_{b,m}^j\right)$. 
Rewritten in terms of the amplitude gain $g(\omega,t)$ defined in Eq.~\eqref{amplitudegain} this leads to
$\alpha \equiv -\big[\frac{\partial}{\partial N}\mathrm{Im}\,g(\omega)\big]/\big[\frac{\partial}{\partial N}\mathrm{Re}\,g(\omega)\big]$
where the relation $\chi(\omega) = 2\varepsilon_{bg}/(i\omega\Gamma) g(\omega)$ between the optical susceptibility and the optical gain $g(\omega)$ was used. The derivative $\partial/\partial N$ in the above definition is, however, ill defined. The contribution to the optical susceptibility, and thus to $\alpha$, is different for each charge carrier transition in the considered laser system. Near-resonant transitions affect mainly the gain, while having little effect on the refractive index, whereas off-resonant transitions mainly contribute to the index change. 
To overcome this problem, we define an $\alpha$-factor via the response of the laser to an optically injected signal, as also done in several experimental studies \cite{LIU01b,LIN11c}. We define
\begin{align}
  \alpha_{inj} \equiv -\frac{\frac{\partial}{\partial K}\mathrm{Im}\,g(\omega)}{\frac{\partial}{\partial K}\mathrm{Re}\,g(\omega)}\,. \label{alphainj}
\end{align}
The above definition overcomes the need to make assumptions about the exact shape of the charge carrier variation $\partial N$ required for evaluating $\alpha$. Instead, the charge carrier variation is determined from the response of the system to the injected signal and thus from the intrinsic system dynamics.
{Other experimental setups yield different charge carrier variations and thus different variations in $g(\omega)$ leading to apparently different $\alpha$. We show below that this issue can be resolved by a full dynamic simulation.}

When injecting a monochromatic optical signal from a master laser into a semiconductor slave  laser, a phenomenon called phase locking emerges, where the phase difference between the electric field inside the cavity and the injected signal becomes constant \cite{STO66}. The slave laser then emits a constant-wave signal with the same frequency as the injected signal. 
Outside the locking range the laser exhibits complex dynamics, including chaos, excitability and multistability \cite{SIM95,OSB09a,WIE02}. The parameter space consists of the injection strength $K$ and the frequency detuning $\Delta\nu_{inj} \equiv (\omega_{inj}-\omega_{fr})/(2\pi)$ between the injected master laser signal and the free-running laser frequency $\omega_{fr}$ of the slave laser.

\begin{figure}[tb]
  \includegraphics[width=8.6cm]{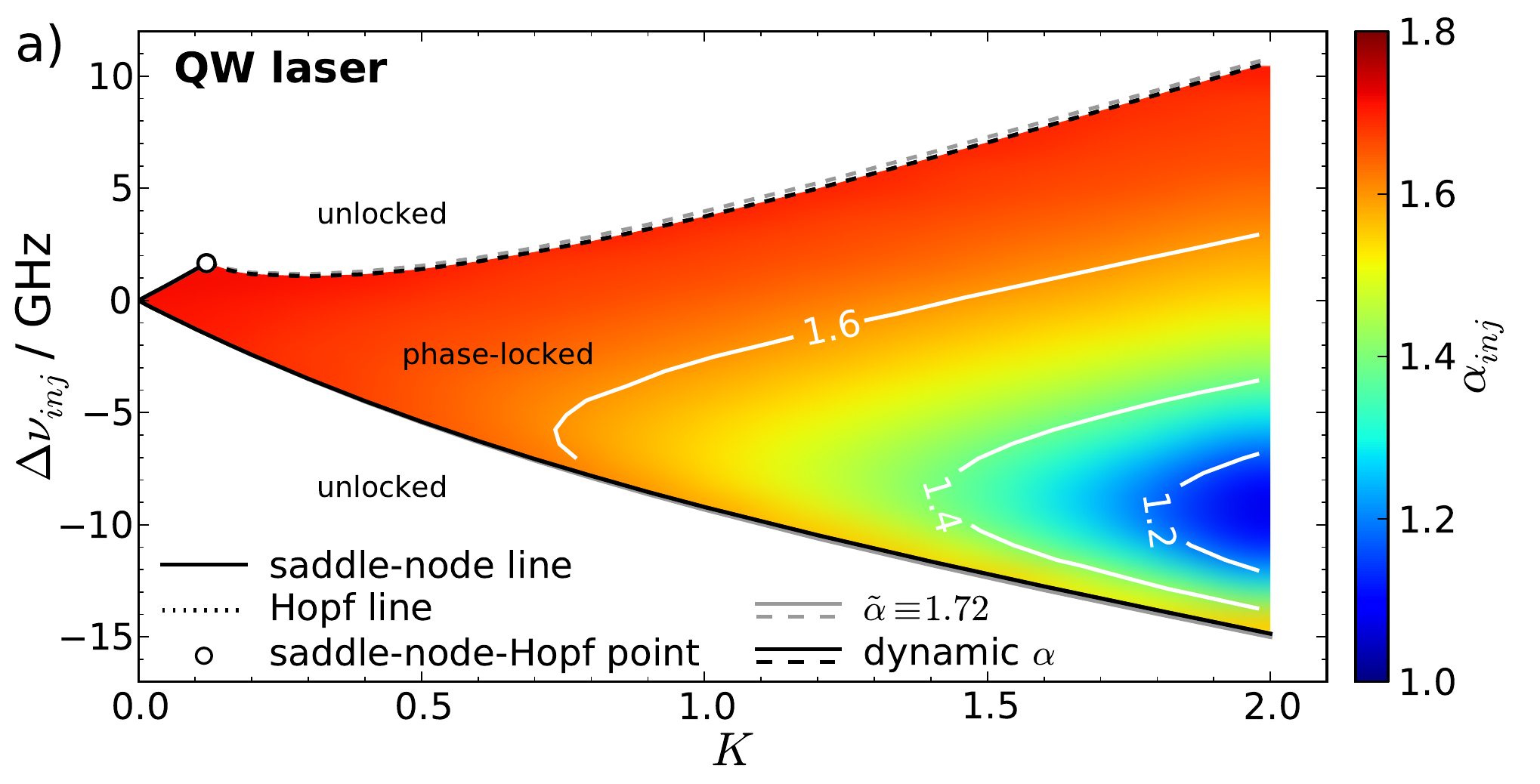}\\
  \includegraphics[width=8.6cm]{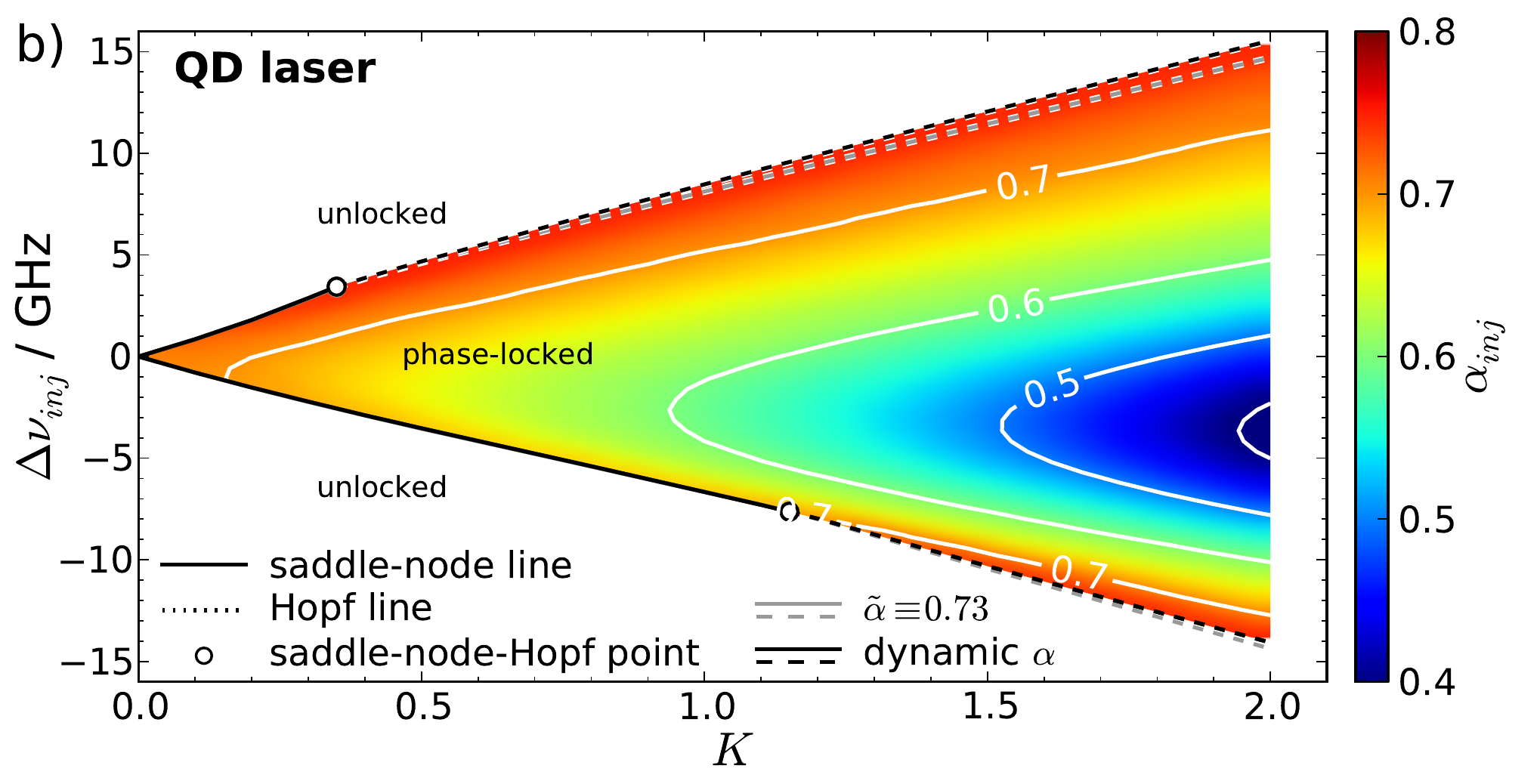}
  \caption{(Color online) Locking tongues in the $(K,\Delta\nu_{inj})$ phase-space for \textbf{(a)} the QW laser, \textbf{(b)} the QD laser. The phase-locked region is bounded by saddle-node (solid lines) and Hopf bifurcations (dashed lines). Black lines correspond to the locking tongue determined for the full model and gray lines to a static effective $\tilde\alpha$. The color code shows the calculated $\alpha$-factor inside the phase-locked region. $J = 2 J_{th}$}
  \label{tongue}
\end{figure}

\begin{figure}[tb]
  \includegraphics[width=8.6cm]{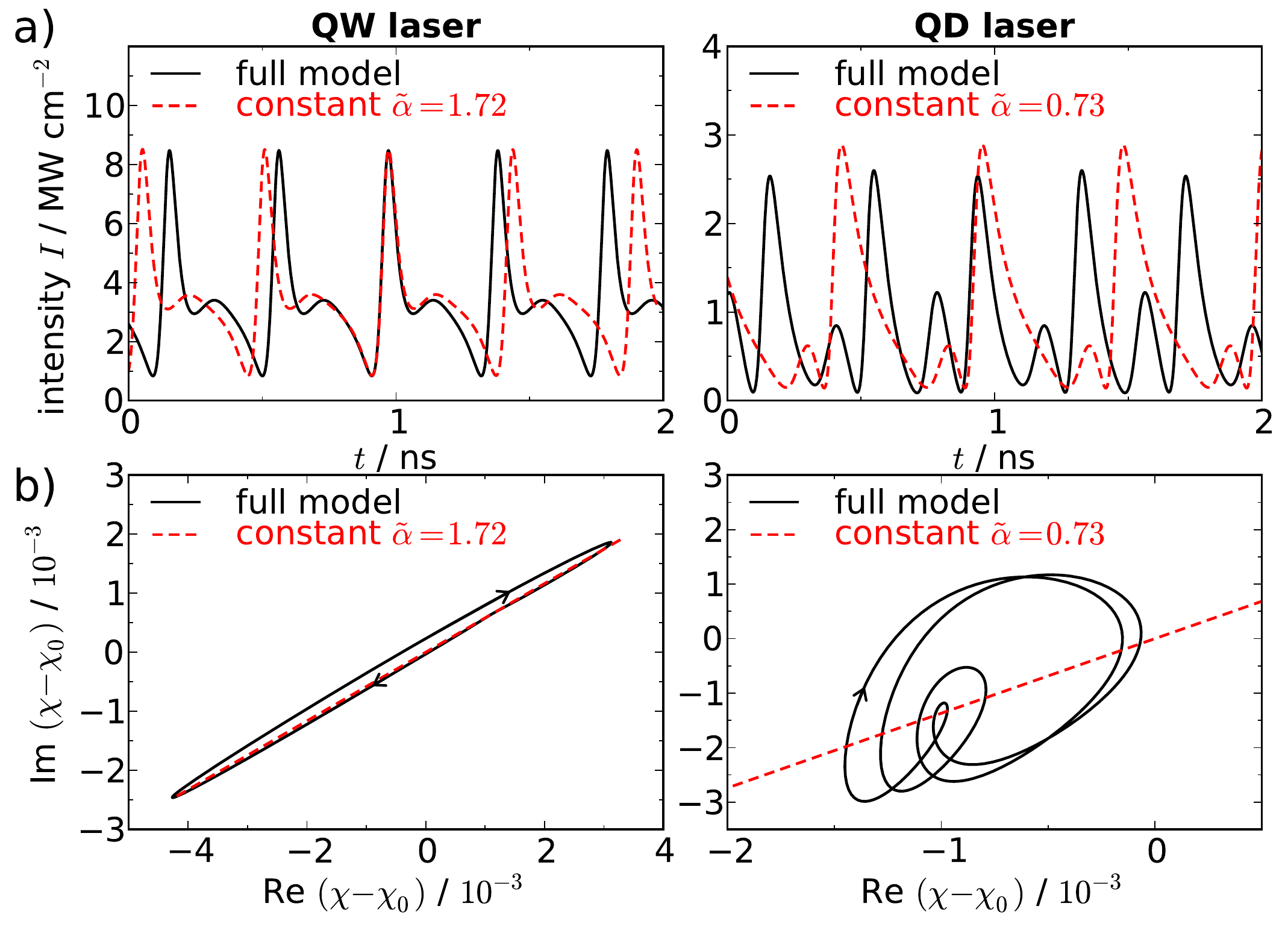}
  \caption{(Color online) Laser dynamics with optical injection outside the locking range. \textbf{(a)} Time-series of the intensity for the QW laser with $K=0.3, \Delta\nu_{inj}=-4.0$GHz (left), and for the QD laser with $K=0.5, \Delta\nu_{inj}=-4.25$GHz (right), with dynamically calculated phase dynamics (black solid line) and with constant effective $\tilde\alpha$-factor (dashed red line). \textbf{(b)} Trajectory in the complex susceptibility plane, shifted by the free-running laser susceptibility $\chi_0$ for the QW (left) and QD laser (right). Same parameters as in \textbf{(a)}. $J = 2 J_{th}$}
  \label{chi}
\end{figure}

The parameter region in the ($K,\Delta\nu_{inj}$)-plane for which phase-locking is possible is shown by the colored region in Fig.~\ref{tongue}(a) and (b) for the QW and QD laser, respectively. It forms Arnold tongues. For low injection strengths, the phase-locked region is limited by saddle-node bifurcations (black solid lines), and becomes limited by Hopf bifurcations (black dashed lines) at higher $K$. 
Since the injection of the optical signal affects the charge carrier distribution of the laser device, we expect the $\alpha$-factor to change when changing the parameters $K,\Delta\nu_{inj}$ due to the change of operating conditions. In order to determine an effective $\alpha$ inside the phase-locked parameter range, we apply Eq.~\eqref{alphainj} at each parameter point, by slightly increasing $K$ and evaluating the changes in gain and index after the transient time. The resulting values for $\alpha_{inj}$ are shown by the color code in Fig.~\ref{tongue}. It can be seen that the effective $\alpha$-factor indeed varies with the injection parameters, showing a considerable decrease inside the locking tongue for both the QD and the QW laser. 
To identify the differences arising from using a constant $\alpha$-factor across the parameter space, we additionally simulate the QD and QW lasers using a gain term corresponding to those used in conventional laser models, by assuming a linear relationship between the refractive index and the gain, given by
$\tilde g(\omega) = ( 1 - i\tilde\alpha )\mathrm{Re}\,g(\omega)$
where the effective $\tilde\alpha$ can be arbitrarily chosen. We use $\tilde\alpha\equiv\alpha_{inj}$ at $K=\Delta\nu_{inj}=0$ as defined in Eq.~\eqref{alphainj}, which should approximate the $\alpha$-factor of the laser under optical injection in the best way possible ($\tilde\alpha=0.73$ in the QD case and $\tilde\alpha=1.72$ in the QW case). Surprisingly, the limits of the locking tongue shown in Fig.~\ref{tongue}  can be very well described by using constant $\alpha$-factors (grey lines) both for the QW and the QD lasers.

\begin{figure}[tb]
  \includegraphics[width=8.5cm]{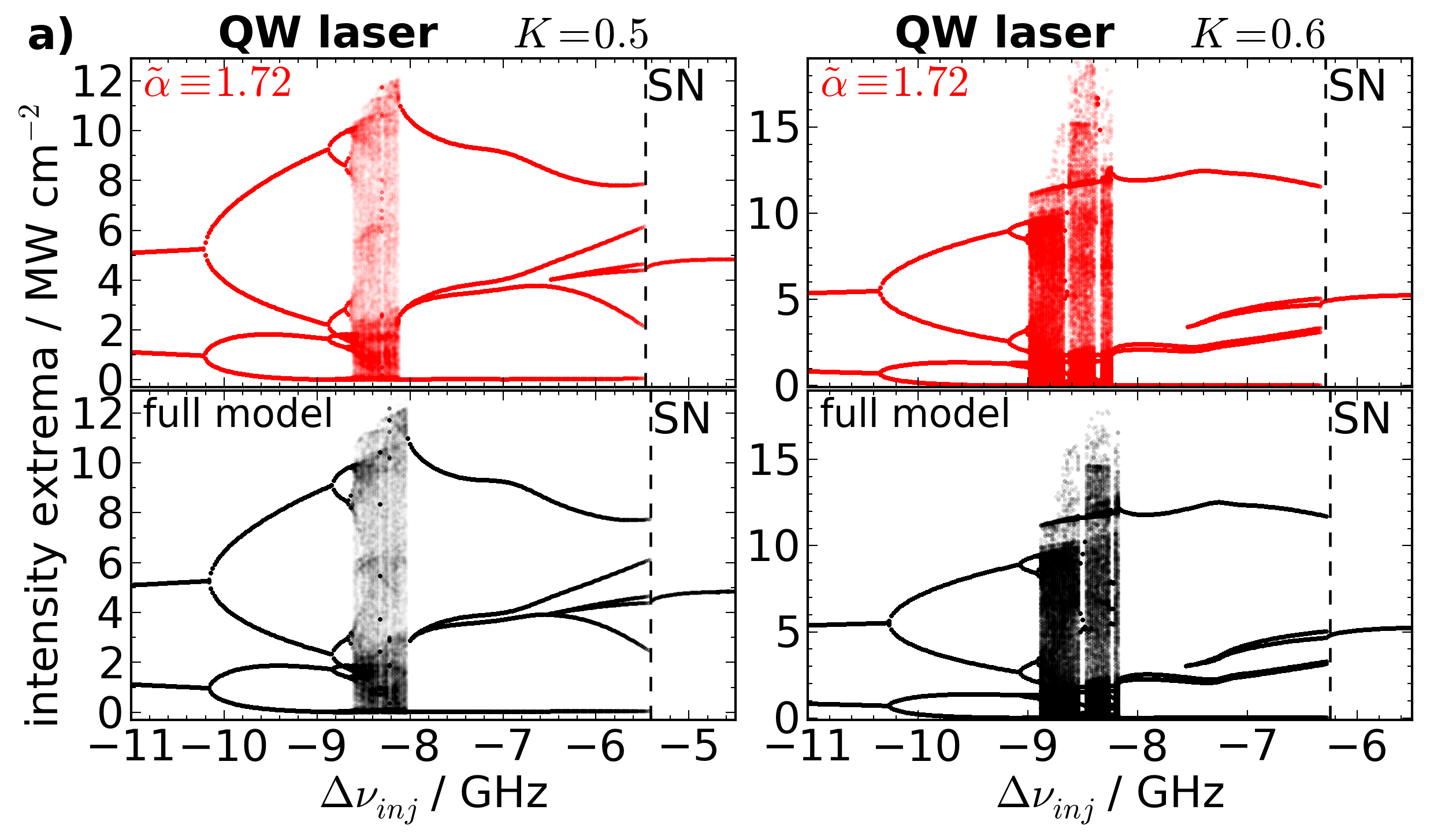}\\
  \includegraphics[width=8.5cm]{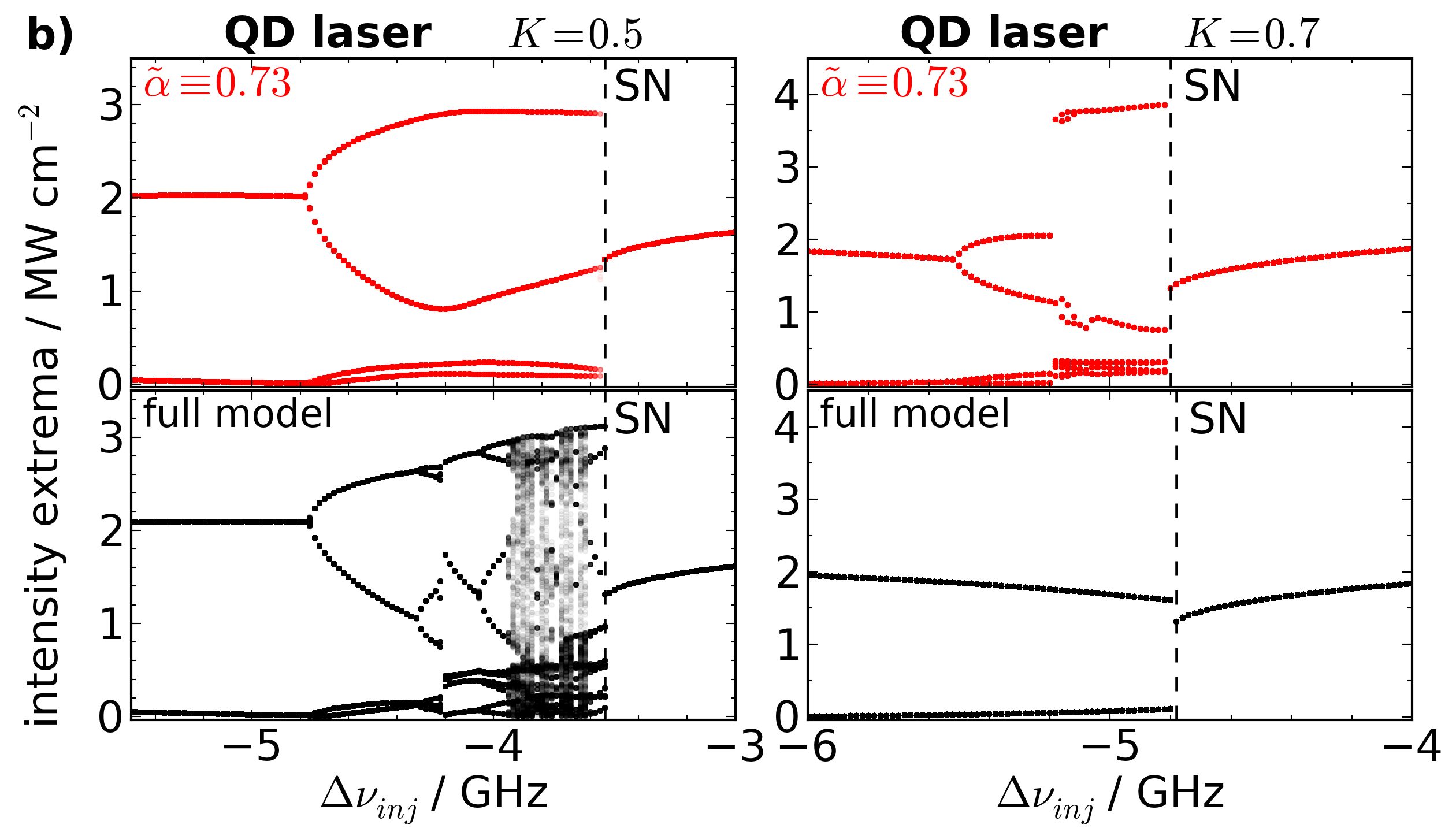}
  \caption{(Color online) Bifurcation diagrams of the output intensity extrema determined by sweeping the injection detuning $\Delta\nu_{inj}$ from the phase-locked detuning range towards lower values for \textbf{(a)} the QW laser at $K=0.5$ (left), $K=0.6$ (right), and \textbf{(b)} the QD laser at $K=0.5$ (left), $K=0.7$ (right). In \textbf{(a)} and \textbf{(b)} top (red) data are obtained using constant $\tilde\alpha$ and bottom (black) data using the full model. 
  The saddle-node (SN) bifurcation limiting the phase-locked regime is shown by the vertical dashed line. $J = 2 J_{th}$}
  \label{bif}
\end{figure}

Outside of the phase-locked parameter range, the optically injected laser exhibits oscillatory intensity pulsations \cite{WIE05,PAU12}. These oscillations occur on a timescale comparable to the charge carrier scattering lifetimes of QD lasers. We therefore expect the dynamics of the gain and refractive index of the QD laser to become important here.  Fig.~\ref{chi} shows a comparison of the laser dynamics outside of the phase-locked region calculated with the full model (black solid lines) and with a constant effective $\tilde\alpha$ (red dashed lines). The time-series of the intensity in Fig.~\ref{chi}(a) reveal qualitatively similar dynamics in the case of the QW laser. Also a near-linear relation between the real and imaginary susceptibility exists as shown in Fig.~\ref{chi}(b), which justifies the assumption of a constant $\tilde\alpha$.
However, for the QD laser case displayed in the right column of Fig.~\ref{chi}(a) the assumption of a constant $\tilde\alpha$-factor leads to qualitatively different dynamics (dashed line showing period-2 oscillation) if compared to simulations with the full model (period-4 oscillations). The reason for this becomes apparent in Fig.~\ref{chi}(b), where the trajectory in the complex susceptibility plane deviates appreciably from the linear relationship given by $\tilde\alpha=0.73$. Here the independent dynamics of resonant and off-resonant states leads to a desynchronization of gain and refractive index in the full model. 
{The time evolution of the optical susceptibility in this case cannot be described by a single $\alpha$-factor.}
{Note that the desynchronization of the real and imaginary part of the susceptibility is observed for all $K$ in the QD laser.}

To shed more light on the differences in the dynamics arising from using a constant $\tilde\alpha$, we calculate bifurcation diagrams of the QD and QW laser dynamics outside of the locking range, using the full model on the one hand, and a constant $\tilde\alpha$ on the other hand (see Fig.~\ref{bif}). By sweeping the injection detuning downwards from the phase-locked parameter range for a constant $K$ and plotting the intensity extrema of the time-series, we can numerically trace bifurcations in the $(K,\Delta\nu_{inj})$ parameter plane. 
As discussed before, the assumption of constant $\tilde\alpha$ does not lead to qualitative differences in the dynamics of the QW laser (see Fig.~\ref{bif}(a)). 
The QD laser, instead, reveals remarkable differences in the bifurcation structure outside the phase-locked region if a constant $\tilde\alpha$ is used (see Fig.~\ref{bif}(b)). At $K=0.5$ the full model exhibits large region of chaotic dynamics, followed by inverse period-doubling bifurcations, below the saddle-node bifurcation. These bifurcations are missing completely if a constant $\tilde\alpha$ is used. 
Also at $K=0.7$ the constant $\tilde\alpha$-factor leads to more complex periodic orbits, while the full model predicts only period-1 oscillations outside the locking range. 
This reveals that the assumption of a linear relation between refractive index and gain is not justified in QD lasers and will eventually lead to an incorrect prediction of the QD laser dynamics. 
Note that in QW lasers the use of $\alpha$ is justified only for sufficiently low injection current and injection strength ($K\lessapprox0.7$). Otherwise non-equilibrium effects become important, invalidating the use of an $\alpha$-factor also for studying complex dynamics in QW lasers.

So far we showed that the use of an $\alpha$-factor in QD lasers is only valid when discussing steady-states. However, even then the $\alpha$-factor needs to be treated with care as it changes for each operation point (i.e., pump current) and experimental setup. Apart from the optical injection, common ways of determining $\alpha$ include the evaluation of frequency modulation and amplitude modulation response \cite{SU05a,MIN97b} and the evaluation of amplified spontaneous emission (ASE) spectra \cite{NEW99a}. 
From simulations of the different experiments for the QW laser, the calculated $\alpha$ is similar for all setups. This is due to very fast relaxation processes coupling the resonant (near band-edge) and off-resonant (higher $k$) states.
As long as non-equilibrium effects can be neglected, i.e., for sufficiently low injection strengths, the charge carrier distribution closely follows a quasi-Fermi distribution. 
{ Therefore a functional relationship of the $k$-resolved carrier distribution on the total charge carrier number exists, and thus the derivative $\partial/\partial N$ is well-defined.}
In QD lasers, however, we find that different experimental setups yield different values for $\alpha$, even at the same operation point \cite{LIN12a} due to the considerably slower carrier scattering between resonant and off-resonant states. 
{ Consequently, measurements of the refractive index dynamics gathered from one experiment should not be used to predict the laser response in a different setup, since the underlying charge carrier dynamics may be different.}


To summarize, {by applying a semi-classical model to evaluate the concept of an $\alpha$-factor in QD and QW lasers, we show that the refractive index dynamics in QD lasers is inaccurately described by $\alpha$.}
We find that in the context of an optical injection setup it is possible to define an effective $\alpha$-factor both for QD and QW lasers when dealing with cw output, but its value varies appreciably with the operating conditions. 
The dynamic response of the QD laser to the injected signal outside of the locking region differs crucially from the dynamics predicted by using a constant $\alpha$, due to the desynchronization of gain and refractive index. Thus the concept of $\alpha$ breaks down. We expect that result to hold also, e.g., in modulation and optical feedback scenarios where the laser emits non-cw output. We conclude that the approximations inferred by introducing an a-factor for the field dynamics are reasonable for QW lasers operated close to equilibrium but are too limiting for modeling complex dynamic scenarios in QD lasers.

This work was supported by DFG within Sfb 787, the U.S. Department of Energy's National Nuclear Security Administration under Contract DE-ACD4-94AL85000, and Sandia's Solid-State Lighting Science Center, an Energy Frontier Research Center (EFRC) funded by the U. S. Department of Energy, Office of Science, Office of Basic Energy Sciences.

 %
 
\end{document}